\documentclass[conference]{IEEEtran}
\makeatletter
\def\ps@headings{%
\def\@oddhead{\mbox{}\scriptsize\rightmark \hfil \thepage}%
\def\@evenhead{\scriptsize\thepage \hfil \leftmark\mbox{}}%
\def\@oddfoot{}%
\def\@evenfoot{}}
\makeatother
\usepackage{array, calc, color, multicol}
\usepackage{mathtools}

\usepackage{graphics, epstopdf, graphicx, epsfig, psfrag, epsf, subfigure}
\usepackage{amssymb, amsfonts, amsmath, times}
\usepackage{relsize}
\usepackage{multicol,balance, verbatim}
\usepackage{textcomp, mathrsfs, stmaryrd, setspace}
\usepackage{booktabs} 
\usepackage{lipsum}
\usepackage{soul}
\usepackage[dvipsnames]{xcolor}
\usepackage{cite}
\usepackage{url}
\usepackage{amsmath,stackengine}
\stackMath

\usepackage{tikz}
\usetikzlibrary{shapes.geometric, arrows, positioning, arrows.meta}
\usepackage{graphicx}

\newcommand{\ie}{{\em i.e., }}

\newtheorem{theorem}{Theorem}

\newtheorem{corollary}[theorem]{Corollary}

\DeclareGraphicsExtensions{.eps, .pdf, .png, .jpg}   % optional

\newcommand{\triot}{{\tt{triOT}}\xspace}

\usepackage{lettrine}
\usepackage[ruled,vlined]{algorithm2e}

\usepackage{amsmath}
\newcommand\numeq[1]%
{\stackrel{\scriptscriptstyle(\mkern-1.5mu#1\mkern-1.5mu)}{=}}

\IEEEoverridecommandlockouts

\begin{document}

\title{Efficient and Privacy-Preserving Binary Dot Product via Multi-Party Computation}

\author{Fatemeh Jafarian Dehkordi$^*$, Elahe Vedadi$^*$, Alireza Feizbakhsh$^*$, Yasaman Keshtkarjahromi$^\dag$, Hulya Seferoglu$^*$ \\
fjafar3@uic.edu, evedad2@uic.edu, afeizb2@uic.edu, yasaman.keshtkarjahromi@seagate.com, hulya@uic.edu \\
$^*$University of Illinois Chicago, $^\dag$Seagate Technology
\thanks{This work was supported in parts by the Army Research Lab (W911NF2420172), Army Research Office (W911NF2410049), the National Science Foundation (CCF-1942878, CNS-2148182, CNS-2112471).}}

\IEEEoverridecommandlockouts
% \makeatletter\def\@IEEEpubidpullup{6.5\baselineskip}\makeatother
%\IEEEpubid{\parbox{\columnwidth}{
%    Network and Distributed System Security (NDSS) Symposium 2024\\
%    26 February - 1 March 2024, San Diego, CA, USA\\
 %   ISBN 1-891562-93-2\\
    %https://dx.doi.org/10.14722/ndss.2024.23xxx\\
%    www.ndss-symposium.org
%}
%\hspace{\columnsep}\makebox[\columnwidth]{}}

\maketitle

\begin{abstract}

Striking a balance between protecting data privacy and enabling collaborative computation is a critical challenge for distributed machine learning. While privacy-preserving techniques for federated learning have been extensively developed, methods for scenarios involving bitwise operations, such as tree-based vertical federated learning (VFL), are still underexplored. Traditional mechanisms, including Shamir's secret sharing and multi-party computation (MPC), are not optimized for bitwise operations over binary data, particularly in settings where each participant holds a different part of the binary vector. This paper addresses the limitations of existing methods by proposing a novel binary multi-party computation (BiMPC) framework. The BiMPC mechanism facilitates privacy-preserving bitwise operations, with a particular focus on dot product computations of binary vectors, ensuring the privacy of each individual bit. The core of BiMPC is a novel approach called Dot Product via Modular Addition (DoMA), which uses regular and modular additions for efficient binary dot product calculation. 
%\textcolor{blue}{To provide privacy, we use random masking in the higher field for linear computations and a three-party oblivious transfer (\triot) protocol for non-linear operations of binary vectors.} 
To ensure privacy, BiMPC uses random masking in a higher field for linear computations and a three-party oblivious transfer (\triot) protocol for non-linear binary operations.   
The privacy guarantees of the BiMPC framework are rigorously analyzed, demonstrating its efficiency and scalability in distributed settings.
% \st{To provide privacy, BiMPC introduces two key protocols: Binary Input Secret Sharing (BISS), an information-theoretic secret sharing mechanism for linear computations of binary data, and Semantically Secure Secret Sharing (SemS$^3$), which provides semantic security for non-linear operations of binary vectors.} 
  %Our approach addresses a significant gap in privacy-preserving computations involving bitwise operations, offering a tailored solution for real-world applications in distributed machine learning systems.
\end{abstract}
\section{Introduction}\label{sec:introduction}

In today's era of big data analytics and distributed machine learning, a critical challenge emerges: how to strike the right balance between protecting data privacy and harnessing distributed data for collaborative computation. The variety of data types and analysis techniques necessitates a differentiated approach to privacy. 
Privacy-preserving mechanisms are well-established for classical federated learning, which typically relies on floating-point operations. Notably, Shamir's secret sharing \cite{shamirSS} and multi-party computation (MPC) protocols based on Shamir’s approach have been widely explored in the literature to offer information-theoretic privacy guarantees for federated learning \cite{secureml1, Zhu2021ImprovedCF, 9333639, 8613446, scalableMPC}. Additionally, techniques like classical homomorphic encryption \cite{homomorphic2, homomorphic, gentry2009fully, van2010fully, fontaine2007survey} and differential privacy \cite{DP_1, DP_2, DP_3, DP_4, DP_5} have been widely used. 

However, privacy-preserving methods remain relatively underdeveloped for computing mechanisms that rely on bitwise operations such as tree-based vertical federated learning (VFL) \cite{qian2025tree}. Some privacy-preserving VFL approaches \cite{VFL2, VFL-Pivot, VFL4, VFL5, VFL6, VFL1} have employed secure hardware, trusted third parties, or cryptographic MPC techniques. However, to the best of our knowledge, no efficient, information-theoretic MPC approach, specifically tailored to the bitwise operations common in VFL, has been proposed. This gap largely stems from the fact that standard oblivious transfer protocols are inherently cryptographic and do not offer information-theoretic security.

%\textcolor{blue}{Yet, to our knowledge, no efficient and specialized information-theoretic MPC solution based on oblivious transfer has been proposed specifically for the bitwise operations common in VFL. This gap is primarily due to the fact that general oblivious transfer protocols are cryptographic and not information-theoretically secure.}

% Yet, to the best of our knowledge, no MPC solution based on Shamir's secret sharing has been developed for VFL mechanisms that require bitwise operations. This gap is primarily due to the fact that Shamir’s secret sharing is not optimized for small finite fields.

In this paper, we focus on bitwise operations of multiple binary vectors, where each vector is held by a different party, and we aim to ensure the privacy of every individual element in all the vectors. While generic protocols like GMW \cite{goldreich2019play} can provide computational security for VFL, they are often prohibitively expensive. 
Similarly, applying Shamir's secret sharing to single-bit data is highly impractical, resulting in significant communication and computation overhead \cite{SS-with-binary-shares}.
%Applying traditional Shamir's secret sharing to protect a single-bit data is extremely inefficient and incurs high communication and computation costs \cite{SS-with-binary-shares}.
Moreover, our problem demands the privacy of each individual bit, meaning that we cannot treat each binary vector of length $n$ as an $n$-bit number and apply Shamir's scheme over a large field. Therefore, there is a need for a tailored secret sharing method that can preserve the privacy of each individual bit in binary secrets.
% Traditional Shamir's secret sharing and multi-party computation (MPC) protocols using Shamir's approach are not applicable in this context. 
% Shamir's secret sharing is an efficient algorithm which encodes a secret data with arbitrary length $l$, such that if the encoded data is shared with any $m \leq 2^l$ participants, for a threshold parameter $z$, (i) the knowledge of any $z$ shares does not reveal any information about the secret, and (ii) any choice of $z+1$ shares fully reveals the secret. This secret sharing scheme requires shares of length $l$, making it optimal for general cases \cite{SS-with-binary-shares}. 
%However, Shamir's scheme cannot be applied directly on single-bit data, because the secret length restricts the number of participants to $ m \leq 2 $, with a maximum of $ z = 1 $ colluding party, which is a trivial scenario.
%

To address this problem, we propose a binary multi-party computing (BiMPC) mechanism. BiMPC is a low complexity multi-party algorithm for privacy-preserving bitwise operations, with a particular focus on dot product computations of binary vectors, ensuring the privacy of each individual bit. BiMPC relies on a novel approach that we name Dot Product via Modular Addition (DoMA) to calculate binary matrix multiplication, which makes BiMPC efficient and scalable. DoMA calculates the dot product of binary vectors via regular and modular additions as further explained in Section \ref{sec:dotpro-modadd}. 
BiMPC is comprised of secure multi-party addition and XOR calculation for linear (regular addition) and non-linear (modular addition) operations of DoMA, respectively. BiMPC uses random masking for secure addition, and XOR operation is performed using a three-party oblivious transfer (\triot) protocol as further detailed in Section \ref{sec:BiMPC}. 
The privacy guarantees of BiMPC are provided in Section \ref{sec:BiMPC}.
%BiMPC is comprised of two novel protocols: BISS and SemS$^3$ for linear (regular addition) and non-linear (modular addition) of DoMA, respectively. BISS is inspired by Shamir's secret sharing, but designed solely for binary vectors as detailed in Section \ref{sec:binarySS}. SemS$^3$ is designed to provide privacy to binary vectors for modular additions as explained in Section \ref{subsec:sems3}. 
%

\section{Related Work}\label{sec:related}

The most widely known secret sharing scheme with information-theoretic security is Shamir’s algorithm \cite{shamirSS}. Although Shamir’s scheme is efficient for sufficiently large finite fields, it is not directly applicable to binary secrets, where field size constraints become a limitation. 
%\textcolor{blue}{Shamir’s scheme is not efficient for binary secrets, due to the large field size constraints that become a limitation.} 
This has motivated several adaptations for small-field secure computation. For example, Chen et al. \cite{chen2006algebraic} propose a generalization using algebraic curves. 
% \st{In this approach, the secret is shared as a vector over the same field such that the shares sum to the secret—contrasting with Shamir’s use of polynomial interpolation. Although this reduces the required field size, it still does not extend to extremely small fields, i.e., for binary vectors.} 
Unlike Shamir’s scheme, which requires the field size to exceed the number of parties, their approach leverages curves of higher genus to support more participants over smaller fields while still preserving strong multiplication properties. Nevertheless, the number of participants is still fundamentally constrained by the field size, so despite this refinement, the approach remains limited and does not scale for very small fields.
% Our proposed method addresses these limitations as it allows privacy-preserving bitwise operations \textcolor{blue}{for an arbitrary number of parties.}

To support MPC over binary vectors, secret sharing schemes based on error-correcting codes have been developed. Notably, Massey \cite{massey1993minimal,massey1995some} introduced a scheme in which a secret is embedded in a random codeword from a binary linear code, while reconstruction relies on the dual code. Building on this, Chen et al. \cite{chen2007secure} extended Massey's approach to support multiplication, making it more suitable for MPC. While this construction is efficient for computation, its security is limited: the access structure (i.e., the subsets of participants who can reconstruct the secret) is fixed by the codebook and does not support arbitrary authorized sets.

Visual cryptography explores the sharing of binary secrets. A foundational method was introduced by Naor and Shamir \cite{naor1995visual}, in which the "transparencies" of a secret binary image are distributed among $n$ participants. The image can be revealed only when at least $k$ participants combine their shares, forming a deterministic $(k,n)$ scheme. However, this approach suffers from pixel expansion, significantly increasing storage requirements for each party. To address this, Wang et al. \cite{WANG20072776} proposed a deterministic scheme that avoids pixel expansion. It uses randomly generated Boolean matrices to produce shares, and the secret image is fully recovered by XORing all the shares. The same work also introduced a probabilistic scheme—similar to random grid methods such as \cite{KafriRG}—in which dark pixels are accurately reconstructed, but bright pixels may occasionally appear as dark. This reconstruction relies partially on human visual perception and thus is unsuitable for scenarios requiring precise recovery. Furthermore, while these methods reduce storage overhead, they lack flexibility in the number of participants required for reconstruction. Deshmukh et al. \cite{deshmukh2019secret} proposed a related approach using a full binary tree structure. The secret is divided into $2^h$ shares, where $h$ is the height of the tree, and recovery requires XORing of all of them. Although efficient in some respects, this scheme also demands the participation of all shareholders, limiting its practicality in flexible threshold scenarios.
Our proposed method addresses these limitations as it allows privacy-preserving bitwise operations efficiently. 
%, in terms of storage and computation.

\section{System Model}\label{sec:system-model}

\emph{Notations.} The set of real numbers, natural numbers, and integers are denoted by $\mathbb{R}$, $\mathbb{N}$, and $\mathbb{Z}$, respectively. The set of numbers from a finite field with $q$ elements is shown by $\mathbb{F}_q$. %\textcolor{blue}{Moreover, integer numbers are denoted by both $a$ and $A$}, and 
We assume that $\Omega_{u}^{v}$ refers to the set of integers between $u$ and $v$, \ie $\Omega_{u}^{v} = \{u, \ldots, v\}$, where $u  \leq v$. 
Matrices and vectors are represented by $\mathbf{A}$ and $\mathbf{a}$, respectively, \ie for $\mathbf{A} \in \mathbb{F}_q^{n \times d}$ and $\mathbf{a} \in \mathbb{F}_q^{n \times 1}$ we have
\begin{align}
\mathbf{A}=\left[ {\begin{array}{ccc}
   a_{1,1}&
   \ldots & a_{1,d}\\   
   \vdots&\ddots&\vdots \\ a_{n,1} &
   \ldots&a_{n,d}
  \end{array} } \right],
\end{align}
for $a_{i,j} \in \mathbb{F}_q$, where $i \in \Omega_1^n$, $j\in \Omega_1^d$, and
\begin{align}
\mathbf{a}^T = (a_1, a_2, \dots, a_n),
\end{align}
for $a_i \in \mathbb{F}_q$. The logical ``AND'' function is shown by the $\land$ symbol, and the ``dot product'' is shown by the $\odot$ symbol. Moreover, concatenation of two vectors $\mathbf{a}$ and $\mathbf{b}$ is represented by $[\mathbf{a}:\mathbf{b}]$, and $\overline{a}$ denotes bitwise complement: $\overline{a} = 1 - a$ for $a \in \{0, 1\}$.

\emph{Problem Statement.} We consider a federated setup with $N$ clients (data owners) and a master device. The clients have data as binary vectors, e.g., client $W_u$ holds binary vector $\mathbf{a}_u$. Clients want the master to calculate the dot product of their binary vectors without revealing the content and length of their vectors, neither to the master nor to any other clients. In other words, the master will learn $y= \mathbf{a}_1 \odot \mathbf{a}_2 \odot \ldots \mathbf{a}_N$ without violating the privacy requirements of the clients.

\emph{Attack Model.} We consider a semi-honest system where all devices follow the defined protocols but are curious about each other's data. We note that the master is a non-colluding device, and no subset of clients colludes as they have no incentive, so no collusion occurs.

%and $z$ out of $N$ clients (for any $z<\frac{N}{2}$) may collude with each other to learn more about the data of non-colluding clients
%\textcolor{blue}{We don't do multiplication on BISS shares, can we have a collusion of $z < N$?}

\emph{Privacy Requirements.} Privacy requirements from the perspective of each client and the master device are defined as follows. 
\begin{itemize}
    \item \textit{Client perspective:} Each client should not learn anything about the private data of other clients from the perspective of information-theoretic security.

    \item \textit{Master perspective:} The master should not learn anything more than the final dot product.
\end{itemize}
The formal definition of privacy requirements is provided in Section \ref{sec:BiMPC}.

\section{Building Blocks of BiMPC Algorithm} \label{sec:buliding-blocks}
In this section, we introduce the building blocks of our proposed BiMPC algorithm, which are (i) Dot Product via Modular Addition (DoMA) and (ii) Three-Party Oblivious Transfer (\triot).

\subsection{Dot Product via Modular Addition (DoMA)}\label{sec:dotpro-modadd}
Let us consider our proposed system model, consisting of $N$ clients and a master device. Our purpose is to compute the dot product of binary vectors owned by clients.  

\begin{theorem}\label{thr:A1&A2&...&Ak}
Consider $\mathbf{a}_1,\;\mathbf{a}_2,\dots,\;\mathbf{a}_l$ as $l$ vectors that are chosen from $\mathbb{F}^{1 \times n}_2$. The AND function of these $l$ vectors can be calculated as the following
\begin{align}
    \mathbf{d} &= \mathbf{a}_1\land \mathbf{a}_2 \land \dots \land \mathbf{a}_l \nonumber \\
    &= \frac{1}{l}\bigg((\mathbf{a}_1+\dots+\mathbf{a}_l)-
\Big((\mathbf{a}_1+\dots+\mathbf{a}_l) \;\text{mod}\; l\Big)\bigg).
\end{align}
\end{theorem}
{\em Proof:} Let us define $\mathbf{d} \in \mathbb{F}^{1 \times n}_2$, {$\mathbf{s} \in \mathbb{F}^{1 \times n}_{l+1}$}, and $\mathbf{m} \in \mathbb{F}^{1 \times n}_l$ as the following:
\begin{align}
    \mathbf{d} := \mathbf{a}_1\land \mathbf{a}_2 \land \dots \land \mathbf{a}_l
     = (d_1,d_2,\dots,d_n),
\end{align}

\begin{align} \label{eq:s}
    \mathbf{s} := \mathbf{a}_1+\dots+\mathbf{a}_l =(s_1,s_2,\dots,s_n),
\end{align}

\begin{align}\label{eq:m}
    \mathbf{m} := (\mathbf{a}_1+\dots+\mathbf{a}_l) \;\text{mod}\; l = (m_1,m_2,\dots,m_n),
\end{align}
where $d_i \in \{0,1\}$, $s_i \in \Omega_0^l$, and $m_i \in \Omega_0^{l-1}$, for $i \in \Omega_1^n$. From the definition of the AND function, we know that $d_i=1$ if and only if the $i^{th}$ element of all of the $\mathbf{a}_1,\dots,\mathbf{a}_l$ vectors is equal to $1$; otherwise, $d_i=0$. Thus, we have
\begin{align}
    d_i = \begin{cases} 
    1 & \text{if~} s_i=l \\
    0 & \text{if~} 0 \leq s_i \leq l-1,
    \end{cases}
\end{align} and 

\begin{align}
     d_i = \begin{cases}
    1 & \text{if~} m_i=0 \\
    0 & 0\leq m_i=s_i \leq l-1
    \end{cases}
\end{align}

Therefore, for $i \in \Omega_1^n$, if the $i^{th}$ element of all of the $\mathbf{a}_1,\dots,\mathbf{a}_l$ vectors is equal to $1$, $d_i=\frac{1}{l}(s_i-m_i)=1$, otherwise $d_i=s_i-m_i=0$, and it concludes the proof. \hfill $\Box$

\begin{corollary}\label{cor:calcA.B}
Consider $\mathbf{a} = (a_1,a_2,\dots,a_n)$ and $\mathbf{b}=(b_1,b_2,\dots,b_n)$ as two binary vectors in $\mathbb{F}^{1 \times n}_2$, where $a_i,\;b_j \in \{0,1\}$ for $i,j \in \Omega_1^n$. The dot product of these two vectors can be calculated as the following:
\begin{align}
y = \mathbf{a} \odot \mathbf{b} = \sum_{i=1}^{n} \frac{1}{2}\bigg((a_i+b_i)-\Big((a_i+b_i)\;\text{mod}\; 2\Big)\bigg).
\end{align}
\end{corollary}
{\em Proof:} The proof can be directly derived from Theorem \ref{thr:A1&A2&...&Ak} and the definition of dot product. \hfill $\Box$

The key advantage of DoMA is its ability to reduce dot product operations to regular and modular additions, significantly lowering communication, computation, and storage costs when applied to binary dot products in MPC.

\subsection{Three-Party Oblivious Transfer} \label{sec:triot}

We use a three-party oblivious transfer (\triot) primitive introduced in our prior work \cite{dehkordi2024privacy}, as a core building block. \triot is inspired by PROXY-OT introduced in \cite{naor1999privacy} providing information-theoretic privacy. In our three-party protocol shown in Fig. \ref{fig:proxy-ot}, the three parties are the selector, sender, and receiver. The selector has a one-bit input $m'$, and the sender has the input keys $\beta_0$ and $\beta_1$. Furthermore, the receiver and the selector generate a random sample $k^m$ using a pseudorandom generator with the same seed, and the receiver and the sender generate random samples $\alpha_0$ and $\alpha_1$ similarly. Here, $m', k^m \in \{0, 1\}$ and $\alpha_j, \beta_j \in \mathbb{F}_q$ for $j = 0, 1$. The goal is for the receiver to obtain $\beta_{m'}$ without learning any other information.\\
First, the selector sends its masked input $m'' = m' \oplus k^m$ to the sender. The sender then forwards the masked labels to the selector. The selector sends $\gamma_{m'} = \beta_{m'} + \alpha_{k^m}$ to the receiver. The receiver can unmask the received key to obtain $\beta_{m'}$. The privacy proof of the \triot protocol is provided in Appendix A of \cite{dehkordi2024privacy}.

\begin{figure}[h]
    \centering
    \resizebox{\columnwidth}{!}{
        \begin{tikzpicture}[
            node distance=1.5cm and 1.5cm,
            >=Stealth,
            every node/.style={font=\footnotesize}
            ]

            % Define positions of nodes
            \node (eval) at (0,0) {\textbf{Receiver}};
            \node[below of=eval, node distance=0.5cm] (eval-inputs) {$\alpha_0,\alpha_1, k^m$};
            \node (garbler) at (4,0) {\textbf{Sender}};
            \node[below of=garbler, node distance=0.5cm] (garbler-inputs) {$\beta_0,\beta_1,\alpha_0,\alpha_1$};
            \node (client) at (8,0) {\textbf{Selector}};
            \node[below of=client, node distance=0.5cm] (client-inputs) {$m',k^m$};

            % Draw arrows for messages
            \draw[->] (client) (8,-1.5) -- (4,-1.5) (garbler) node[midway, above] {$m'' = m' \oplus k^m$};
            \draw[->] (garbler) (4,-2.25) -- (8,-2.25) (client) node[midway, above] {$\gamma_0 = \beta_0 + \alpha_{m''}$} node[midway, below] {$\gamma_1 = \beta_1 + \alpha_{1 - m''}$};
            \draw[->] (client) (8,-3.25) -- (0,-3.25) (eval) node[midway, above] {$\gamma_{m'} = \beta_{m'} + \alpha_{k^m}$};

            % Draw a box around the entire protocol
            \draw[thick] (-1,0.5) rectangle (9,-3.75);

        \end{tikzpicture}
    }
    \caption{\triot protocol.}
    \label{fig:proxy-ot}
    % \vspace{-10pt}
\end{figure}
\section{Binary Multi-Party Computation (BiMPC)} \label{sec:BiMPC}

In the previous section, we explained the foundational components of the BiMPC algorithm. This section explores their integration within an MPC framework, offering privacy proofs for operations conducted on binary secrets. For clarity, we first describe BiMPC for $N = 2$ and then extend it to the general case $N>2$.

\subsection{BiMPC Algorithm for $N = 2$ Clients} \label{sec:bimpc_for_2}

In this system configuration, client $W_1$ holds a binary input vector $\mathbf{a} \in \mathbb{F}_2^{1 \times n}$, client $W_2$ holds a binary input vector $\mathbf{b} \in \mathbb{F}_2^{1 \times n}$, and a master device computes the dot product $y=\mathbf{a} \odot \mathbf{b}$ in a privacy preserving manner.\footnote{While we articulate our algorithm for computing the dot product of two binary vectors, it can be extended to calculate the multiplication of any two binary matrices. This generalization holds as each matrix multiplication can be decomposed into dot products between the rows of the first matrix and the columns of the second matrix.} 
% Consider a system configuration constructed of $N$ clients and a master device. Within this arrangement, two specific clients, referred to as clients $W_1$ and $W_2$, possess binary vectors $\mathbf{a} \in \mathbb{F}_2^{1 \times n}$ and $\mathbf{b} \in \mathbb{F}_2^{1 \times n}$, respectively.
% The objective is to compute the dot product of these two vectors
% %\footnote{We note that it is straightforward to generalize  BiMPC to more than two binary vectors.}
% , denoted as $y=\mathbf{a} \odot \mathbf{b}$, ensuring privacy preservation. 
% \st{in collaboration with the remaining $N-2$ clients and the master device, ensuring privacy preservation.} 
To reduce the complexity of computation, communication, and storage, we employ the DoMA algorithm to calculate the dot product. Beginning with the definition of the dot product of two vectors, we first compute $\mathbf{d} = \mathbf{a} \land \mathbf{b}$ and then derive the final result as $y=\mathbf{a} \odot \mathbf{b} = \sum_{i=1}^n \mathbf{d}[i]$. The DoMA algorithm provides the following expression for $\mathbf{d}$ as

\begin{align}
\mathbf{d} = \mathbf{a} \land \mathbf{b} = \frac{1}{2} \bigg((\mathbf{a} + \mathbf{b}) - \Big((\mathbf{a} + \mathbf{b}) \;\text{mod}\; 2\Big)\bigg), 
\end{align}  which holds component-wise for binary vectors. Let $\mathbf{s} := (\mathbf{a}+\mathbf{b}) \bmod q$ and
$\mathbf{m} := \mathbf{a} \oplus \mathbf{b}$. Therefore, the dot product can be written as
\begin{align}
    y = \mathbf{a} \odot \mathbf{b} = \frac{1}{2} \sum_{i = 0}^{n-1} \Bigl(\mathbf{s}[i] - \mathbf{m}[i]\Bigr).
\end{align}
This reduces the task to two sub-problems: (i) secure addition of shares (to obtain $\sum_i \mathbf{s}[i]$) and (ii) secure XOR computation (to obtain $\sum_i \mathbf{m}[i]$). To ensure the correctness of the final calculation, we carry out all operations (except XOR operations) in a prime field $\mathbb{F}_q$ with $q > 2n$, so that $\bigl(2^{-1}\bmod q\bigr)$ exists and the reconstructed field element equals the integer dot product.\footnote{Noting that the calculation is correct, since $\mathbf{a}[i], \mathbf{b}[i] \in \{0,1\}$ and $q > 2n$, we have $\mathbf{s}[i] \in \{0,1,2\}$ and $\sum_i \mathbf{s}[i] < q$ and the reduction modulo $q$ leaves $\sum_i \mathbf{s}[i]$ unchanged.} The BiMPC algorithm then proceeds as follows:

% To compute these values using the BiMPC algorithm, considering the presence of $z$ colluding clients among $N$ clients \textcolor{violet}{(where $z < \frac{N}{2}$)}, we proceed with the following steps:

\textbf{Step 1 - Addition: Masking data in the higher field.} In the first step, clients $W_1$ and $W_2$ mask their secret data $\mathbf{a} \in \mathbb{F}_2^{1 \times n}$ and $\mathbf{b} \in \mathbb{F}_2^{1 \times n}$ using their keys, respectively, and calculate $\mathbf{s}'_1$ and $\mathbf{s}'_2$ as follows:
\begin{align}
\mathbf{s'}_1 = \sum_{i=1}^{n} \Bigl(\mathbf{a}[i] + \mathbf{k}_1^s[i]\Bigr),
\end{align}
\begin{align}
\mathbf{s'}_2 = \sum_{i=1}^{n} \Bigl(\mathbf{b}[i] + \mathbf{k}_2^s[i]\Bigr),
\end{align} where $\mathbf{k}_1^{s}, \mathbf{k}_2^{s} \in \mathbb{F}_q^{1 \times n}$ are uniformly random vectors.

\textbf{Step 2 - Addition: Padding masked data and sharing with the master.} In this step, the clients pad their masked data with binary vectors of length $n'$ to mask the true length of the input vectors $n$ from the master. 
\begin{align}
\mathbf{s''}_1 = [\mathbf{s'}_1 : \mathbf{p}^s_1],
\end{align}
\begin{align}
\mathbf{s''}_2 = [\mathbf{s'}_2 : \mathbf{p}^s_2],
\end{align} where $\mathbf{p}_1^{s}, \mathbf{p}_2^{s} \in \mathbb{F}_q^{1 \times n'}$ are uniformly random vectors. Then, they send their masked shares $\mathbf{s''}_1$ and $\mathbf{s''}_2$ to the master device.

\textbf{Step 3 - Addition: Calculating the addition of secret shares received from clients.} In this step, the master sums the shares received from client $W_1$ and $W_2$ as follows:
\begin{align}
\mathbf{s''} = (\mathbf{s''}_1 + \mathbf{s''}_2) \;\text{mod}\; q.
\end{align}
% Then, they share the secret share results $\mathbf{a'} \in \mathbb{F}_2^{1 \times 2(n+n')}$ and $\mathbf{b'} \in \mathbb{F}_2^{1 \times 2(n+n')}$ with the master device.

\textbf{Step 4 - XOR: Masking the input data in the binary field.} In this step, client $W_2$ masks its input vector $\mathbf{b}$ by a random key $\mathbf{k}_2^m \in \mathbb{F}_2^{1 \times n}$ and sends the resulting value $\mathbf{b} \oplus \mathbf{k}_2^m$ to client $W_1$.

\textbf{Step 5 - XOR: Calculating the modular 2 addition of the
received secret share.} Client $W_1$ combines its input vector $\mathbf{a}$ with the received share from client $W_2$, $\mathbf{b} \oplus \mathbf{k}_2^m$, and computes the masked value of the modular 2 addition of the input vectors:
\begin{align}
\mathbf{m'} = \mathbf{a} \oplus \mathbf{b} \oplus \mathbf{k}_2^m.
\end{align}

\textbf{Step 6 - XOR: Preparation for the \triot protocol.} Client $W_1$ acts as the \emph{selector}, client $W_2$ as the \emph{sender}, and the master as the \emph{receiver}.
From Step~5, $W_1$ holds the selector bits $\mathbf{m}'=\mathbf{a}\oplus\mathbf{b}\oplus\mathbf{k}_2^m \in \mathbb{F}_2^{1\times n}$.
Client $W_2$ samples a fresh mask $\mathbf{k}\in\mathbb{F}_q^{1\times n}$ and forms the two sender messages
\begin{align}
    \boldsymbol{\beta}_0 = \mathbf{k}_2^m + \mathbf{k},
\end{align} 
\begin{align}
    \boldsymbol{\beta}_1 = \overline{\mathbf{k}_2^m} + \mathbf{k}.
\end{align}

% (ii) Client $W_1$ concatenates a random binary vector $\mathbf{p} \in \mathbb{F}_2^{1 \times n'}$ to its result from previous step, $\mathbf{m'}$, and computes $[\mathbf{m'} : \mathbf{p}]$.

\textbf{Step 7 - XOR: Execution of \triot.} In this step, the three parties jointly execute the \triot protocol. Client $W_1$ inputs the vector $\mathbf{m'}$ (from Step 5), while client $W_2$ inputs the keys $\boldsymbol{\beta}_0$ and $\boldsymbol{\beta}_1$ (from Step 6). The protocol is performed element-wise on these inputs according to Fig. \ref{fig:proxy-ot} and the master obtains $\mathbf{m} + \mathbf{k}$.

\textbf{Step 8 - XOR: Masking the length of the input vectors.} To conceal the true length of the input vectors, clients $W_1$ and $W_2$ agree on a random padding vector $\mathbf{p}^m \in \mathbb{F}_q^{1 \times n'}$, which client $W_1$ sends to the master. From the master's perspective, this operation is indistinguishable from the real XOR-message pattern. Consequently, the master holds the final concatenated vector
\begin{align}
    \mathbf{m''} = [\mathbf{m} + \mathbf{k} : \mathbf{p}^m],
\end{align} whose true length is now hidden.

% \begin{align}
% \mathbf{F}_{\mathbf{m'}}(x) & = \text{BISS}(\mathbf{m'})=\mathbf{P}^{{\mathbf{m'}}}_0+\dots+\mathbf{P}^{{\mathbf{m'}}}_{z-1}x^{z-1}+\mathbf{m'}x^z,
% \end{align} where $\mathbf{P}^{{\mathbf{m'}}}_i \in \mathbb{F}_q^{1 \times 2(n+n')}$ are uniformly random vectors. Afterwards, it shares $\mathbf{F}_{\mathbf{m'}}(u) \in \mathbb{F}_q^{1 \times 2(n+n')}$ with client $W_u$ for $u \in \Omega_1^N$.

\textbf{Step 9: Calculating the addition of the random keys.} In this step, clients $W_1$ and $W_2$ compute $k'_1$ and $k'_2$, the addition of the keys they used in the previous steps, respectively, as follows:
\begin{align}
k'_1 = \sum_{i=1}^{n} \mathbf{k}_1^s[i] + \sum_{j=1}^{n'} \Bigl(\mathbf{p}^s_1[j] - \mathbf{p}^m[j]\Bigr)
\end{align}
\begin{align}
k'_2 = \sum_{i=1}^{n} \Bigl(\mathbf{k}_2^s[i] - \mathbf{k}[i]\Bigr) + \sum_{j=1}^{n'} \mathbf{p}^s_2[j]
\end{align}

Finally, both clients send their resulting scalar values to the master.

\textbf{Step 10: Calculating the final dot product.} In this step, the master uses all the components it has and computes the final dot product $y = \mathbf{a} \odot \mathbf{b}$ as follows:
\begin{align}
    \mathbf{d''} = \mathbf{s''} - \mathbf{m''},
\end{align}
\begin{align}
    y = \frac{1}{2} \biggl(\sum_{i=1}^{n+n'} \mathbf{d''}[i] - k'_1 - k'_2 \biggr) \,\text{mod}\, q.
\end{align}

% Then, it calculates the summation of all elements of $\mathbf{F}_{\mathbf{d}}(u)$ as
% %\st{plus a uniformly random number $k \in \mathbb{F}_q$, which is known by all clients}
% \begin{align}
% \mathbf{F}_{y}(u) = \sum_{i=1}^n \mathbf{F}_{\mathbf{d}}(u)[i].
% \end{align}
% Finally, client $W_u$ sends $\mathbf{F}_{y}(u) \in \mathbb{F}_q$ to the master. 

% \textbf{Step 10: Decoding the final result.} In this step the master waits for at least \st{$z$} \textcolor{blue}{$(z+1)$} results, i.e., $\mathbf{F}_{y}(u)$ from \st{$z$} $(z+1)$ clients and \st{encodes} decodes $y = \mathbf{a} \odot \mathbf{b}$ using the Lagrange interpolation rule.

\subsection{BiMPC Algorithm for $N > 2$ Clients}
The BiMPC framework generalizes from the two-client case to accommodate an arbitrary number of clients, $N>2$, through an iterative application of the core protocol. In this setting, each of the $N$ clients holds a binary vector $\mathbf{a}_u$, and the objective is for the master to securely compute the final dot product $y = \mathbf{a}_1 \odot  \mathbf{a}_2 \odot \ldots \odot \mathbf{a}_N$.

The computation unfolds sequentially. Initially, the protocol is executed on the first two vectors, $\mathbf{a}_1$ and $\mathbf{a}_2$, to compute their logical AND. The master receives the output of this step only in a securely masked form, which prevents it from learning the intermediate result. This masked value then serves as the first operand in the next iteration, where its logical AND is computed with the third vector, $\mathbf{a}_3$. This procedure is repeated for all remaining vectors up to $\mathbf{a}_N$.

While the secure summation component of the protocol remains unchanged throughout this process, the secure XOR computation requires a minor adaptation. Specifically, the roles within the \triot protocol are reassigned to account for the master's state. The master, now holding the masked intermediate result from the previous step, takes on the role of the \emph{sender} in the \triot protocol. This is a key departure from the two-client case, where a client with a plaintext vector would act as a sender. This iterative design ensures that the master learns nothing more than the final dot product. The specifics of how the parties interact in this $N$-client setting are provided in the extended version of this paper.

\subsection{Privacy Analysis} The formal privacy requirements for the BiMPC algorithm from the perspective of each client and the master device are defined as the following:

\textit{Client perspective:} Each client should not learn anything about the private data of the other clients from the perspective of information-theoretic security. For each client $W_u$:

\begin{align}\label{privacy-for-wu}
    \tilde{H}\big(\mathbf{a}, \mathbf{b}|\text{view}_{W_u}\big)
&=\tilde{H}\big(\mathbf{a}, \mathbf{b}\big),
\end{align} 

where $\tilde{H}$ denotes the Shannon entropy and view$_{W_u}$ denotes the view of client $W_u$ from the messages it receives from other parties.
% and $\Nset_c$ is a subset of $\Nset = \Omega_1^N$ satisfying $|\Nset_c| \leq z$. 
    
\textit{Master perspective}: The master should not learn anything more than the final result $y=\mathbf{a} \odot \mathbf{b}$, \ie
\begin{align}\label{privacy-for-master}
    \tilde{H}\big(\mathbf{a},\mathbf{b}|y, \text{view}_{M}\big)= \tilde{H}\big(\mathbf{a}, \mathbf{b}|y\big),
\end{align} 
where view$_{M}$ denotes the view of the master from the messages it receives from the clients.

%     \begin{align}\label{privacy-for-master}
% \tilde{H}\big(\mathbf{a},\mathbf{b}|y,\underset{u \in \Nset}{\bigcup}\mathbf{F}_{y}(u), \mathbf{a'},\mathbf{b'} \big)= \tilde{H}\big(\mathbf{a}, \mathbf{b}|y\big).
%     \end{align}
%where \textcolor{blue}{To Do for Elahe: Add some explanations about the above equations...} 
%is the data received from $W_n$ by the master node. 

\begin{theorem}
Our proposed BiMPC algorithm satisfies privacy requirements from the perspective of both the clients and the master, presented in Equations (\ref{privacy-for-wu}) and (\ref{privacy-for-master}). 
\end{theorem}

{\em Proof:} The proof of this theorem directly follows from the secrecy of random masking and \triot.
% the results of Theorem \ref{thr:I(A;tilde_FA)=0} and Theorem \ref{thr:SemS3-privacy}.
%\input{BiMPC_for_DT}
%\input{theoretical_analysis}
%\input{perforemance-evaluation}
\section{Conclusion}\label{sec:conclusion}

This paper introduced the Binary Multi-Party Computation (BiMPC) framework, addressing the privacy-preserving challenges in distributed machine learning involving binary data. BiMPC enables efficient and secure dot product computations on binary vectors, leveraging the novel Dot Product via Modular Addition (DoMA) approach. 
% \st{We also proposed two protocols—Binary Input Secret Sharing (BISS) and Semantically Secure Secret Sharing (SemS$^3$)—to preserve the privacy of binary vectors.} 
BiMPC ensures private computation on binary vectors by random masking and \triot protocol. The proposed solution is scalable and provides strong privacy guarantees for distributed machine learning methods that rely on bitwise operations, such as tree-based VFL applications.

\bibliographystyle{IEEEtran}

\bibliography{refs}

%\input{appendix A}

% \section{Discussion} 

% BISS($A$) = $\mathbf{F}_{\mathbf{A}}(x) = \mathbf{P}_0+\mathbf{P}_1x+\mathbf{P}_2x^2+\dots+\mathbf{P}_{z-1}x^{z-1}+\mathbf{A}x^z$.

% Shamir($A$) = $\mathbf{\tilde{F}}_{\mathbf{A}}(x) = A + \mathbf{P}_1x+\mathbf{P}_2x^2+\dots+\mathbf{P}_{z-1}x^{z-1}+\mathbf{P_z}x^z$.

% Assume $z=2$. When $x=2$, Shamir will be $\mathbf{\tilde{F}}_{\mathbf{A}}(2) = A + 2 P_1 + 4 P_2$. If $2P_1 + 4P_2 < q$, where $q$ is the field size, $\mod_2(\mathbf{\tilde{F}}_{\mathbf{A}}(2))$ reveals information. Assume $P_1=P_2=P$ for simplicity, then if $P < \frac{q}{6}$,  $\mod_2(\mathbf{\tilde{F}}_{\mathbf{A}}(2))$ reveals information. It does not provide perfect privacy anymore. 

% Now, assume the same example where $z=2$ and $x=2$, BISS will be $\mathbf{{F}}_{\mathbf{A}}(2) = P_0 + 2 P_1 + 4 A$. In this case, $\mod_2(\mathbf{{F}}_{\mathbf{A}}(2))$ does not reveal any information as $P_0 + 2P_1$ can be odd or even with equal probability. 

\end{document}